\documentclass[twocolumn,aps, prd]{revtex4}%
\usepackage{graphicx}
\usepackage{dcolumn}
\usepackage{bm}
\usepackage{amsmath}%
\usepackage{amsfonts}%
\usepackage{amssymb}

\begin{document}

\title{On the Davies-Unruh effect in a wide range of temperatures
}

\author{Carlos. E. Navia}
\affiliation{Instituto de F\'{\i}sica, Universidade Federal Fluminense, 24210-346,
Niter\'{o}i, RJ, Brazil}

\date{\today}
\begin{abstract}
The Debye model of the specific heat of solid at low temperatures is incorporate in the Entropic Gravity Theory (EGT). Rather of a smooth surface, the holographic screen is considered as an oscillating elastic membrane, with a continuous range of frequencies, that cuts off at a maximum (Debye) temperature, $T_D$.  We show that at low temperatures $T < T_D$, the conservation of the equivalence principle in EGT requires a modification of the Davies-Unruh effect. 
While the maintenance of Davies-Unruh effect requires a violation of the equivalence principle. These two possibilities are equivalents, because both can emulate the same quantity of dark matter.  However,  in both cases, the central mechanism is the Davies-Unruh effect, this seems to indicate that the modification of the Davies-Unruh effect emulates dark matter which in turn can be see as a violation of the equivalence principle.  This scenario is promising to explain why MOND theory works at very low temperatures (accelerations) regime, i. e., the galaxies sector. We also show that in the intermediate region, for temperatures slightly lower or slightly higher than Debye temperature, EGT predicts the mass-temperature relation of hot X-ray galaxy clusters.

\end{abstract}


\maketitle

\section{Introduction}

The Davies-Unruh effect (DHE) \cite{davi75,unru76}, essentially predict that in an accelerated frame of reference; a vacuum state may seen as a thermal bath of photons 
with a black boddy spectrum at a temperature T, the main point of 
the DUE is that this temperature is proportional to the 
acceleration of the frame.

 The connection between thermodynamic and gravity began in the 70s with Bekenstein \cite{beke73} and Hawking \cite{hawk74}, researching the nature of black holes. In 1995 Jacobson \cite{jaco95}shown a thermodynamic description of gravity obtaining the Einstein's equations.

According to Padmanabhan \cite{padm15}, the association between gravity and entropy leads in a natural way to describes gravity as an emergent phenomenon, and a formalism of gravity as 
a entropic force is derived by Verlinde \cite{verl11} in 2010. The dependence of information on surface area, rather than volume (Holographic principle)
\cite{hoof93}, it is one of the key  of black hole thermodynamic theory, as well as in EGT.

On the other hand, the Tully-Fisher relation \cite{tull77} is an empirical result, very well established for spiral galaxies. This relation is hard to be obtained from Newtonian gravity, at least if only the visible mass of the galaxy is considered. The output for this impasse was postulating the presence of a galactic 
halo of dark matter. Nowadays the empirical roots of the missing mass problem
goes from the flat rotation curves of galaxies, cluster of galaxies, 
gravitational lensing, large scale structure, and it is needed to describe 
the spectrum of the cosmic microwave background radiation (CMB).

The nature of dark matter is unknown. But 
the most widely accepted hypothesis is that dark matter is composed of weakly interacting massive particles (WIMPs) that interact only through gravity and the weak force \cite{behn11}.
However, so far, there is no direct evidence of wimps or other dark particles such as the axions, and only upper limits were reported \cite{aker16,tan16}.
In addition, so far, there is no evidence of a new physics beyond 
standard model, in the  Large Hadron Collider (LHC) data \cite{gibn16}.
 There is no evidence for SUSY (super-symmetric particles) \cite{jung96}, where the ``neutralino'' is a kind of natural candidate for a dark matter particle. 

In the 80s, Milgrom \cite{milg83,milg83b}, proposed 
a modification in the Newtonian law of gravity as
solution to the missing mass problem, 
without dark matter.  Rather than a scientific theory MOND is considered only as an empirical model by most of the scientific community, because predicts a 
violation of the equivalence principle. 
However, MOND  has been very well successful 
to describe the galaxies dynamics \cite{mcga11,fama12,sand90,krou10}. Indeed, MOND 
predicted the Tully-Fisher relation.

The connection between entropic gravity and MOND is not new, there are some literature on this topic such as reported in \cite{man10,verl16}. 
 In this Letter we gives emphasis to the formalism of Debye model \cite{deby12} of the specific heat of solid at low temperatures, incorporated to the entropic gravity. 
 In section II, the basis of EGT within the Debye formalism is presented and we defined the Debye temperature, $T_D$, in EGT.
 Section III, is devoted to a analysis of the inertia at low temperatures ($T < T_D$).
In the intermediate region, i.e., temperatures close to $T_D$, EGT seems to indicate that is the galaxy clusters region, this topic is discussed in section IV, and the section V  is devoted for our conclusions.
 
 \section{Gravity at low temperatures}

In 1912, Debye \cite{deby12} developed a theory to explain the heat capacity of solid as low temperatures. He assumed that the vibration of the atoms of the lattice of a solid,
follows a continuous range of frequencies, such as an elastic structure, that cuts off at a maximum frequency, $\omega_D$. 
In this theory each solid has a specific temperature, called as Debye temperature, $T_D=\hbar \omega_D/k_B$. The Debye model correctly predicts the low temperature dependence of the heat capacity of solid and coincides with the  approaching the Dulong-Petit law at high temperatures. 
 
In EGT an holographic screen is the closed area where is stored the information of the surrounding matter enclosed by the screen. The information is codified in N bits and is considered as the freedom degrees of the system. The holographic screens coincides with the Newtonian equipotential surfaces. 

A central argument of the holographic principle is consider that each bit of information on the screen carries an energy $1/2k_BT$ and the number of bits N on the screen surface is proportional to the area of the screen, and expressed as $N=(c^3/G \hbar) A$, where G is Newton gravitational constant. Taking into account the equipartition of energy principle, the specific potential energy on the screen can be written as
\begin{equation}
U=\frac{1}{2}N k_B T \mathcal{D}_1\left(\frac{T_D}{T}\right).
\end{equation}
Following the analogy with Debye model, we have substituted $k_BT$, for $k_BT\;\mathcal{D}_1(T_D/T)$. The main difference with the Debye theory is in that the third Debye function, $\mathcal{D}_3(x)$ was replaced by the first Debye function, $\mathcal{D}_1(x)$, because the information bits located on the screen have only a vibrational state along of the gradient of Newton potential and it is assumes that the vibrations follows a continuous range of frequencies. $\mathcal{D}_1$ is defined as
\begin{equation}
\mathcal{D}_1\left(\frac{T_D}{T}\right)= \left(\frac{T}{T_D}\right) \int_0^{T_D/T} \frac{x}{e^x-1}dx,
\end{equation}
The shape of Debye function reflects the Bose-Einstein statistic formula, used in its derivation. 
  
  IF M represents the all mass  enclosed by the screen surface, the specific potential energy can be written as $U=Mc^2$,
  and considering that the entropy variation, $\Delta S$ of the screen, 
happens when a particle of mass m is at a distance $\Delta X$
(close to the Compton wave length). The Bekestain entropy variation can be expressed as
\begin{equation}
\Delta S=2\pi k_B \frac{mc}{\hbar} \Delta X.
\end{equation} 
these relations, allows to obtain the entropic force defined as $F=T \frac{\Delta S}{\Delta x}$.
The more simple case is for a spherical screen of radius R, ($A=4\pi \;R^2$), and the acceleration of the mass m is
\begin{equation}
a\mathcal{D}_1\left(\frac{T_D}{T}\right)= \frac{GM}{R^2}.
\end{equation}

\begin{figure}
\vspace*{-1.0cm}
\hspace*{-0.0cm}
\centering
\includegraphics[width=9.0cm]{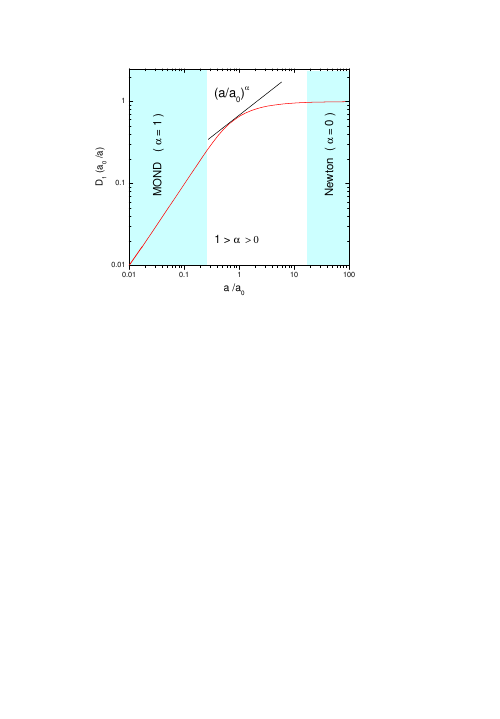}
\vspace*{-8.0cm}
\caption{
Debye first function, as a function of the acceleration and parametrized as a power function. 
}
\label{fig1}
\end{figure} 

Considering that $T/T_D \propto a/a_0$ (see below), the Debye function can be parametrized by a power function of 
type $(a/a_0)^{\alpha}$, as shows in Fig. 1 and  Eq.4 becomes
\begin{equation}
a\left(\frac{a}{a_0}\right)^{\alpha}= \frac{GM}{R^2}.
\end{equation}
The two asymptotically  cases are: (a) $\alpha=0$ and means high temperature regime $T/T_D \gg 1$ and  the Eq.5 coincides with the Newtonian gravity theory, and (b)
 $\alpha=1$ means low temperature regime $T/T_D \ll 1$ and Eq.5 reproduce the Tully Fisher relation $M\propto v^4$ observed in the galaxies  
dynamic and plotted as $\log M=4 v +\log (1/Ga_0)$ \cite{mcga11}.  The slope, 4,  fall precisely with
that observed in  galaxies, whereas the normalization require $a_0=10^{-10}ms^{-2}$ \cite{mcga11}. The acceleration $a_0$ is the Milgrom acceleration parameter \cite{milg83} and in this limit EGT coincides with the MOND theory. 
In this limit the Debye first function (Eq.2) can be related as $\mathcal D_1(x)=\pi^2/x$ with 
$x=T_D/T=2\pi c k_B T_D /(\hbar a)$ and lead to $a_0=12c k_B T_D /(\pi \hbar)$.

Finally the intermediate region, $0 < \alpha < 1$, EGT seems to indicate that is the galaxy cluster region, see section IV.

\section{Inertia at low temperatures}

The starting point for development of the general theory of relativity 
was the equivalence principle, it is also 
valid in the Newtonian gravity. 
There are strong evidences indicating that the equivalence principle holds in all experiments at Earth\cite{murp13}.

We starting the analysis, taken into account the Bekenstein entropy variation expressed in the Eq. 3, and the entropic force concept
$F=T\Delta S/\Delta x$ to obtain
\begin{equation}
F=m_i\;a=2\pi\left[k_BT \;\mathcal{D}_1(T_D/T)\right] m_gc/\hbar.
\end{equation}
According to the Debye framework, the quantity, $k_BT$ was substituted by
$k_BT \;\mathcal{D}_1(T_D/T)$. 
In Eq.6, F in the left side represent the force of inertia, that can be expressed as $F=m_ia$, where $m_i$ is the inertia mass, while the mass, m, in the right side of the equation, represent the mass of the particle, at a distance equal  to Compton wavelength of the holographic screen, when the entropic force emerges, then it is linked with the gravitational mass, $m_g$. Here we have two possibility:

(a) If the equivalence principle holds for all temperature regions, we have $m_i=m_g$ and Eq. 6 becomes
\begin{equation}
a=\frac{2\pi c}{\hbar} k_BT\;\mathcal{D}_1(T_D/T).
\end{equation}
This equation we called as generalized DUE, at high temperatures, $T_D/T \ll 1$, $\mathcal{D}_1(x)\simeq 1$, and coincides with the conventional DUE. Fig. 2 shows a comparison between the generalized and conventional DUE. The discrepancy for $T/T_D \ll 1$ can emulate dark matter. Considering that $m\propto  1/a$ and $m_d \propto 1/a_d$ for dark matter and calling as $a_m$ and $a$ the accelerations according to the modified and conventional DUE, we have $1/a_d=1/a_m - 1/a$ and multiplying this last expression by a, we have
\begin{equation}
\frac{m_d}{m}=\frac{1}{\mathcal{D}_1(T_D/T)}-1.
\end{equation}
For high temperatures $T/T_D \gg 1$, $\mathcal{D}_1(T_D/T)=1$ and $m_d=0$.

\begin{figure}
\vspace*{-0.5cm}
\hspace*{-0.0cm}
\includegraphics[width=7.0cm]{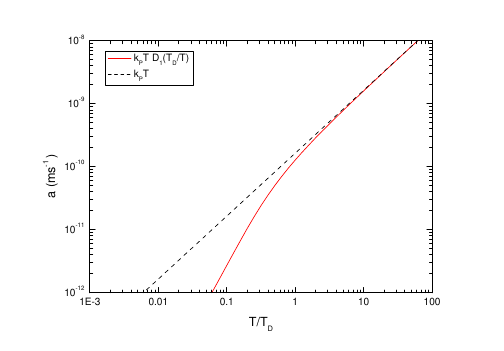}
\vspace*{-0.5cm}
\caption{
Comparison between the conventional DUE, dotted black line and the generalized DUE
effect, solid red curve. 
}
\label{fig2}
\end{figure} 

(b) If the equivalence principle is violated $m_i \neq m_g$ and  keeping the DUE without modification Eq.6 becomes 
\begin{equation}
\frac{m_i}{m_g}=\mathcal{D}_1(T_D/T).
\end{equation}
Taking in account Eq. 5 $\mathcal{D}_1(T_D/T)=(a/a_0)^{\alpha}$, the ratio $m_i/m_g$ is plotted in Fig. 3 as function of acceleration.

\begin{figure}
\vspace*{-1.5cm}
\hspace*{-0.0cm}
\centering
\includegraphics[width=9.0cm]{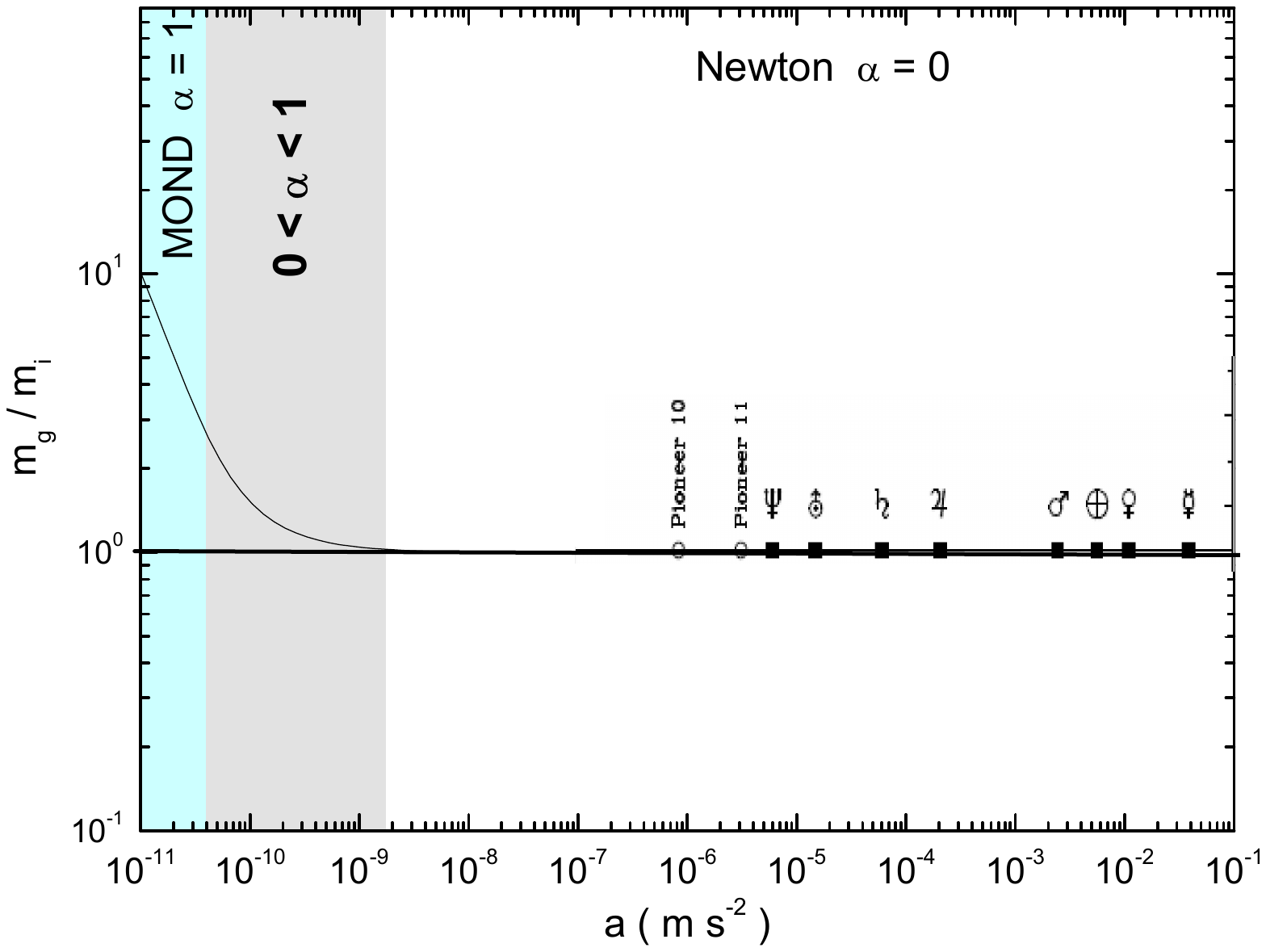}
\vspace*{-6.0cm}
\caption{
The ratio between the gravitational and inertial masses, as a function of the acceleration. The  figure extended to solar-system scales
(each planet is labelled). 
}
\label{fig3}
\end{figure} 

Following the Fig.3, we can see that the ratio $m_g/m_i$ for low accelerations
increases as the acceleration decreases, the extreme case (strong violation)  coincides with the MOND theory prediction. 
While, for high accelerations (temperatures), $a/a_0 \gg 1$, and it includes all solar system, the equivalence principle remains valid.

The fact that $m_i \neq m_g$ 
 for motions in low temperatures regimen, can emulate dark matter. If we called as $m_d$ the dark matter mass, it can be obtained as $m_d=m_g-m_i$ and taking into account Eq.8, we obtain
\begin{equation}
\frac{m_d}{m_i}=\frac{1}{\mathcal{D}_1(T_D/T)}-1;
\end{equation}
This expression coincides with Eq. 8, and means that we have a duality.
However, in both cases, the central mechanism is the DUE. This suggested that the modification of DUE is more fundamental in EGT and the emulated dark matter is seen as an apparent violation of the equivalence principle.

\section{Galaxy clusters}

It is well known also that the MOND theory has its ``Achilles' heel''. The galaxy cluster, seems to indicate that still is necessary a residual mass. In most cases the MOND critics largely use this to reject MOND.
Indeed, the residual mass required by MOND was supply with an exotic neutrino, the ``sterile neutrino'', considered as  promising candidates to hot dark matter \cite{angu08,angu11}. 
So far,  there is no direct evidence of these neutrinos  \cite{aart16,adam16}.

On the other hand, in the central part of clusters the observed acceleration is usually slightly larger than $a_0$ \cite{bell03}. This clearly shows the limitations of MOND in cluster analysis.
However, there is not this limitation in EGT and means that the temperature of the holographic screen that surrounding clusters enclosed by the screen has a temperature slightly larger than the Debye temperature. This means that the clusters analysis requires
$0 < \alpha < 1$.

The relationship between various galaxy cluster mass estimators and X-ray gas temperature agree to within 40\% \cite{horn99}.
Virial theorem mass estimates based on cluster galaxy velocity
dispersions seem to be accurately  related to the X-ray temperature as 
$M \propto T^{\delta}$ with $\delta=3/2$ \cite{horn99}. This results are consistent with that predicted by simulations \cite{evra96}. However, when is combined several independent observation, i.e., a wide range of temperatures of galaxy clusters, seems to indicate a steeper 
$\delta \lesssim 2$ index \cite{fama12}. Even so, in all cases, the greatest discrepancy is in normalization, with differences around 40\% to 50\%.

An analysis of galaxy cluster on the basis of entropic gravity is presented in \cite{verl16}. However, in this section we present an alternative straightforward analysis on galaxy clusters, on the basis of EGT within the Debye formalism. 

\begin{figure}
\vspace*{-3.0cm}
\hspace*{-0.0cm}
\centering
\includegraphics[width=11.0cm, height=9cm,angle=90]{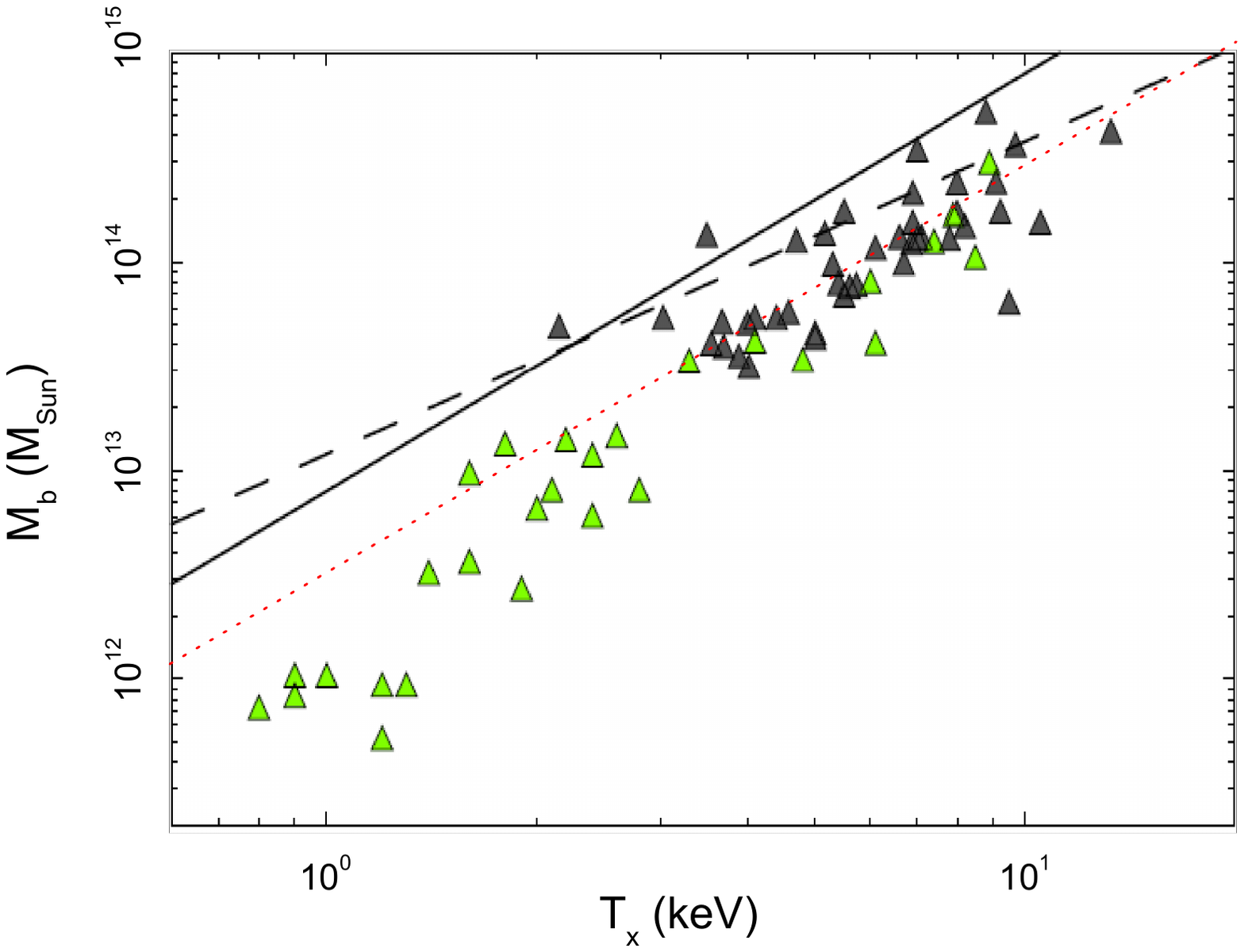}
\vspace*{-3.5cm}
\caption{
The mass  X-ray temperature relation for galaxy clusters (gray triangles \cite{reip02,sand08})
and groups of galaxies (green triangles \cite{angu08}). The dashed line indicates the expected in 
$\Lambda$CDM \cite{evra96}. The solid line indicates the prediction of MOND \cite{fama12}, ($\alpha=1$),
 and the red dot line is the prediction of EGT within the Debye formalism to $\alpha=0.96$. 
 }
\label{fig4}
\end{figure} 

Under the assumption of spherical symmetry and following the Eq. 5, the asymptotic $(r \rightarrow \infty)$, allow us calculate the mass of cluster as
\begin{equation}
M(r\rightarrow \infty)= \frac{r^2}{G} a (\frac{a}{a_0})^{\alpha}.
\end{equation}
The acceleration in the gravitational potential of the cluster is related by 
$a=C_s^2d\ln \rho_x/dr$, with $C_s^2=k_bT_x/\nu m_p$. The density distribution $\rho_x$ is well described by the ``$\beta -model''$, whose asymptotic expression is $\rho_x \sim r^{-3\beta}$, where $\beta$ has a typical value of $\sim 2/3$ \cite{gerb92}. Under these conditions the Eq. 11 becomes
\begin{equation}
M(r\rightarrow \infty)= \frac{r^{1-\alpha}}{G a_0^{\alpha}}C_s^{2(1+\alpha)}(-3\beta)^{1+\alpha} .
\end{equation}

For $\alpha=1$ we have the MOND prediction to the mass-Xray temperature relation of clusters, 
expressed as $M \propto T_X^2$. This relation is represented by the black solid line in Fig. 4 \cite{fama12}, where the mass function of an X-ray flux of several samples of galaxy clusters are
plotted.  We can see that data is closer with  MOND's predicted slope than that previsioned by standard $\Lambda$CDM and expressed as $M \propto T_X^{3/2}$ \cite{evra96} and represented by the dashed black line in Fig. 4, and it is better than MOND only at high X-ray temperatures.
But the previsioned MOND's normalization is around two times greater than observed.

The red dots line in Fig. 4, represent the prediction of Eq. 12, for 
$\alpha=0.96$ and expressed as $M \propto T_X^{\delta}$ with $\delta=\alpha +1=1.96$. The normalization for this case differs from MOND normalization by a factor 
$a_0^{1-\alpha}\;=0.40$, for $\alpha=0.96$ and $a_0 \sim 10^{-10}ms^{-1}$.

\section{Conclusions}

EGT within the Debye formalism can emulate dark matter of two equivalent ways, The first case requires modification of the DUE in order to maintain the equivalence principle and the second one requires modification of equivalence principle in order to maintain the DUE.  In both cases the DUE is the fundamental mechanism. This means that
the modification of the DUE  emulate dark matter which in turn emulates a violation of the equivalence principle. This scheme is promising, because recently results \cite{mcga16}, on the SPARC database of galaxies \cite{lell16}, seems indicate a challenge to the dark matter hypothesis. 

The EGT within the Debye formalism is also promising in the analysis of galaxy clusters. However, there is also a second dark entity to consider, the dark energy, responsible of the accelerating expansion of Universe. The dark energy in EGT is discussed in \cite{eass11,verl16}. Even so, there is more a complication, the increase  in a factor of 10 of the number of supernovae IA, in relation to the first analyses, seems to indicate that the accelerated expansion signal is only marginal \cite{niel16}. 

We believe that as more information is gathered, we will have more conditions to test the EGT within the Debye formalism, in other complex systems.

\acknowledgments

This work is supported by the National Council for Research (CNPq) of
Brazil, under Grant 306605/2009-0.

\end{document}